\documentclass[]{emulateapj}

\begin{document}

\slugcomment{Accepted in ApJ Letters}

\newcommand{\kms}{km s$^{-1}$}
\newcommand{\msun}{$M_{\odot}$}
\newcommand{\rsun}{$R_{\odot}$}
\newcommand{\vsini}{$v\sin{i}$}
\newcommand{\teff}{$T_{\rm eff}$}
\newcommand{\logg}{$\log{g}$}
\newcommand{\tarr}{$t_{\rm arr}$}

\title{ The Nature of Hypervelocity Stars and the Time Between Their Formation and Ejection }

\author{Warren R.\ Brown$^1$,
	Judith G.\ Cohen$^2$,
	Margaret J.\ Geller$^1$, and
	Scott J.\ Kenyon$^1$}

\affil{ $^1$Smithsonian Astrophysical Observatory, 60 Garden St, Cambridge, MA 02138\\
	$^2$Palomar Observatory, Mail Stop 249-17, California Institute of Technology, Pasadena, CA 91125 }


\shorttitle{ The Nature of Hypervelocity Stars }
\shortauthors{Brown et al.}

\begin{abstract}

	We obtain Keck HIRES spectroscopy of HVS5, one of the fastest unbound stars
in the Milky Way halo.  We show that HVS5 is a $3.62\pm0.11$~\msun\ main sequence B
star at a distance of $50\pm5$~kpc.  The difference between its age and its flight
time from the Galactic center is $105\pm18$(stat)$\pm$30(sys)~Myr; flight times from
locations elsewhere in the Galactic disk are similar.  This $10^8$~yr `arrival time'
between formation and ejection is difficult to reconcile with any ejection scenario
involving massive stars that live for only $10^7$~yr.  For comparison, we derive
arrival times of $10^7$~yr for two unbound runaway B stars, consistent with their
disk origin where ejection results from a supernova in a binary system or dynamical
interactions between massive stars in a dense star cluster.  For HVS5, ejection
during the first $10^7$~yr of its lifetime is ruled out at the 3-$\sigma$ level.  
Together with the $10^8$~yr arrival times inferred for three other well-studied
hypervelocity stars, these results are consistent with a Galactic center origin for
the HVSs.  If the HVSs were indeed ejected by the central black hole, then the
Galactic center was forming stars $\simeq$200~Myr ago, and the progenitors of the
HVSs took $\simeq$100~Myr to enter the black hole's loss cone.

\end{abstract}


\section{INTRODUCTION}

	\citet{hills88} first predicted unbound ``hypervelocity'' stars (HVSs) as
the inevitable consequence of 3-body interactions close to the tidal radius of a
massive black hole.  There is overwhelming evidence for a $4\times10^6$~\msun\
central black hole in the Milky Way \citep{ghez08, gillessen09}.  Theorists expect
that the black hole ejects $\sim$$10^{-4}$ HVSs yr$^{-1}$ \citep[e.g.][]{perets07},
which means there are thousands of HVSs in the outer halo.  \citet{brown05}
discovered the first HVS, a luminous B-type star traveling twice the Galactic escape
velocity at a distance of $\simeq$100 kpc, and \citet{brown12b} have subsequently
discovered 15 more unbound B-type stars in their targeted HVS survey.  Establishing
the evolutionary state of the HVSs is important for establishing their ages,
distances, and flight times.  We define the difference between a HVS's age and its
flight time as the `arrival time' (\tarr), the time between its formation and
ejection.  In this paper we derive \tarr\ for both HVSs and unbound runaway stars.

	The arrival time provides a useful discriminant between proposed ejection 
mechanisms.  If HVSs are ejected in three-body interactions with the Milky Way's 
central black hole \citep{hills88}, then the arrival times reflect the timescale for 
HVSs to achieve orbits that interact with the central black hole.  For HVSs formed 
in the central region of the Galaxy, we expect \tarr$=0.1$--1~Gyr \citep{merritt04, 
wang04}.  On the other hand, in both mechanisms for ejecting runaway stars from the 
Galactic disk -- a supernova in a binary system or a dynamical interaction among 
massive stars in a dense star cluster -- a maximum \tarr$\approx10$~Myr is set by 
the main sequence lifetime of $\gtrsim$10~\msun\ stars.  Thus, measuring \tarr\ for 
an ensemble of HVSs should distinguish between a Galactic center or Galactic disk 
origin.

	The evolutionary state of most known HVSs \citep{brown12b} is ambiguous 
because their effective temperatures and surface gravities are consistent with both 
old, evolved stars (blue horizontal branch stars) and short-lived main sequence 
stars.  Thus we must turn to other measures to establish their nature.  Metallicity 
is one possibility; we expect that recently formed stars should have solar or 
super-solar metallicities.  Metallicity is inconclusive, however, given the observed 
metallicity distribution function of stars in the Milky Way.

	Projected stellar rotation \vsini\ is a better discriminant between evolved
stars and main sequence stars.  Blue horizontal branch stars have evolved through
the giant branch phase and have median \vsini\ $=9$~\kms; the most extreme blue
horizontal branch star rotates at 40~\kms\ \citep{behr03}.  Late B-type main
sequence stars, on the other hand, have median \vsini\ $=150$~\kms; the most extreme
objects rotate at $\ge$350~\kms\ \citep{abt02, huang06a}.  \citet{lockmann08} argue
that HVSs may be spun-up by a binary black hole ejection, but there is presently no
evidence for a binary black hole in the Galactic center. Close binaries that produce
HVSs and runaways in the Milky Way may exhibit slower stellar rotation because of
tidal synchronization; \citet{hansen07} predicts that late B-type HVSs ejected by
the Hills mechanism should have \vsini\ $=70-90$ \kms.  In any case, fast rotation
is the signature of a main sequence star.

	Of the B-type HVSs discovered to date, only HVS3, HVS7, and HVS8 have been
studied with high-resolution spectroscopy.  In all cases they are main sequence B
stars with 55$<$\vsini$<$260 \kms\ \citep{edelmann05, przybilla08, bonanos08,
lopezmorales08, przybilla08b}.  Moderate-dispersion spectroscopy of HVS1 suggests it
has \vsini\ $=190$~\kms\ \citep{heber08b}, another short-lived B star.

	Here, we describe high resolution spectroscopy of HVS5, a $g=17.9$ mag star 
located at declination +67\arcdeg\ accessible only with Keck HIRES.  HVS5 is a 
rapidly rotating 3.6~\msun\ main sequence B star.  The difference between its age 
and its flight time from the Milky Way is $105\pm18$(stat)$\pm$30(sys)~Myr, 
inconsistent with ejection models involving massive stars.

	In Section 2 we describe the observations and stellar atmosphere analysis.  
In Section 3 we discuss the arrival times for the HVSs and unbound disk runaways.  
We conclude in Section 4.

\section{DATA}

\subsection{Observations}

	We observed HVS5 using the HIRES spectrograph \citep{vogt94} at the 10~m
Keck~1 telescope, obtaining four 1800 sec exposures the night of 2012 Jan 29, and
five more the following night.  Both nights were clear with 0.7--0.8 arcsec seeing.  
The red HIRES collimator was used in an instrument configuration that gave spectral
coverage from 3920 to 8350~\AA.  A 1.1~arcsec slit gives a spectral resolution of
$R=$34,000, with 6.7 pixels per spectral resolution element.  There are small gaps
between the three CCDs that form the detector mosaic, sometimes resulting in the
loss of all or part of a single echelle order.

	We also observed eight B stars selected from \citet{abt02} that span a wide 
range in luminosity class (I, III, and V) and a large range in projected rotational 
velocity ($5<$~\vsini~$<285$~\kms).  The stars are HR1328, HR1333, HR1399, 
HR1419, HR1420, HR1462, HR1573, HR1595, and HR1640.  We used high S/N spectra of 
these eight B stars to provide a comparison for HVS5 and to validate our analysis 
below.

	We used the pipeline package {\sc makee}\footnote{{\sc~makee} was developed
by T.A.\ Barlow specifically for reduction of Keck HIRES data.  It is freely
available from the Keck HIRES home page \url{www2.keck.hawaii.edu/inst/hires}.} to
remove the instrumental signature, extract a one dimensional spectrum for each
echelle order, and calibrate the wavelength scale from Th-A arc spectra taken at the
beginning and end of each night.  Each exposure of HVS5 was individually processed
through {\sc makee}, and the results summed.  With a total integration of 4.5 hours,
we achieved a $S/N$ ratio of 70 per spectral resolution element at 4500~\AA.

\begin{figure}		
 \plotone{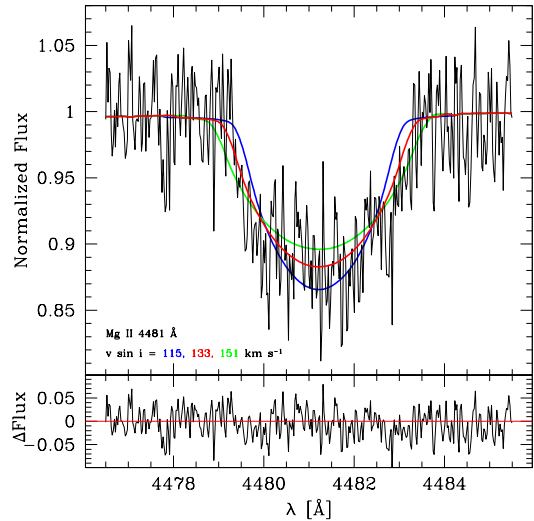}
 \caption{ \label{fig:vsini}
	Observed Mg {\sc ii} $\lambda$4481 \AA\ line (upper panel) and its residuals 
(lower panel) compared to the best-fit \vsini$=133\pm7$~\kms\ model (red line).  For 
reference, the green and blue lines are the $\pm$3-$\sigma$ models.}
 \end{figure}

\subsection{Spectral Analysis}

	Visual inspection reveals that HVS5 is a fast rotator.  The full width at 
half maximum of the Mg {\sc ii} $\lambda$4481 line compared with the \citet{abt02} B 
stars suggests a projected rotation velocity of $\simeq$130~\kms.  We turn to the 
most recent ATLAS9 model atmosphere grids \citep{castelli04, castelli97} to perform 
a more quantitative analysis.  We use the program {\sc spectrum} \citep{gray94} to 
calculate synthetic spectra under the assumption of plane-parallel atmosphere and 
local thermodynamic equilibrium.  We adopt macro- and microturbulunce velocities of 
0 and 2~\kms, respectively.  All synthetic spectra are smoothed to a resolution of 
$R=34,000$ and re-sampled with spline interpolation to match the sampling of the 
observed spectra.

	Our approach is to analyze the spectral lines on an order-by-order basis.  
We normalize the continuum, calculate the $\chi^2$ of each synthetic model against
the data, and then fit the resulting distribution of $\chi^2$ to derive the
best-fitting parameters.  Uncertainty estimates are provided by a standard
$\Delta\chi^2$ offset to the minimum $\chi^2$ \citep{press92}.  Our final values are
the weighted means and error-on-the-means of the parameters measured from lines in
different orders.

	We begin by using all of the spectral lines to solve for the heliocentric 
radial velocity.  The best-fit $+552\pm3$~\kms\ velocity is in perfect agreement 
with the $+553\pm9$~\kms\ velocity measured from medium-resolution spectroscopy at 
the MMT \citep{brown12b}.  The constancy in velocity is consistent with HVS5 being a 
single star, as one expects for the product of a binary disruption.  The radial 
velocity corresponds to a minimum velocity of +663~\kms\ in the Galactic rest frame 
\citep[see][]{brown12b}.

	Next, we measure projected rotation using Mg {\sc ii} $\lambda$4481 \AA, the 
strongest metal line in the spectrum (see Figure~\ref{fig:vsini}).  The best-fit 
\vsini\ is $133\pm7$~\kms.  For comparison, Figure~\ref{fig:vsini} plots the 
$\pm3$-$\sigma$ values as well as the residuals to the best-fit \vsini.  The 
observed \vsini\ is consistent with the median \vsini\ of comparable B-type main 
sequence stars \citep{abt02, huang06a}.

\begin{figure}		
 \plotone{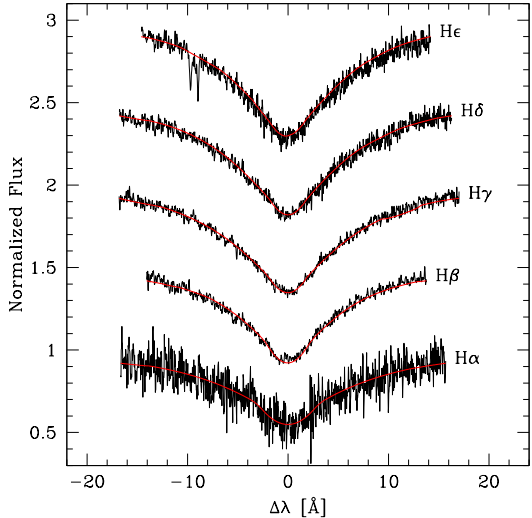}
 \caption{ \label{fig:tplot}
	Observed hydrogen Balmer lines compared to the best-fit model (red line).  
The temperature- and surface gravity-sensitive lines give best-fit values of 
\teff$=12,000\pm350$ K and \logg$=3.89\pm0.13$.}
 \end{figure}

	Given the observed \vsini, we measure effective temperature and surface
gravity by fitting the widths and depths of the \teff- and \logg-sensitive hydrogen
Balmer lines.  We note that the model hydrogen lines are computed using the
D.~Peterson routine adopted by {\sc synthe} \citep{kurucz93}, which includes Stark
and resonance broadening and fine structure in the cores.  The best-fit values are
\teff$=12,000\pm350$ K and \logg$=3.89\pm0.13$; Figure~\ref{fig:tplot} compares the
best-fit model with the data.

	Finally, we attempt to constrain the metallicity.  Because of the large 
\vsini, Fe lines are faint continuum fluctuations and are thus too weak to provide 
significant constraint.  Si {\sc ii} lines at $\lambda$4128, $\lambda$4131, and 
$\lambda$5056 \AA\ are stronger and yield a best-fit Si abundance of 
[M/H]=$-0.4\pm0.5$.  The Mg {\sc ii} $\lambda$4481 \AA\ line (see 
Figure~\ref{fig:vsini}), on the other hand, yields a best-fit Mg abundance of 
[M/H]=$+0.3\pm0.5$.  Given the large uncertainties, we conclude that HVS5 is 
consistent with solar abundance.

\begin{figure}		
 \plotone{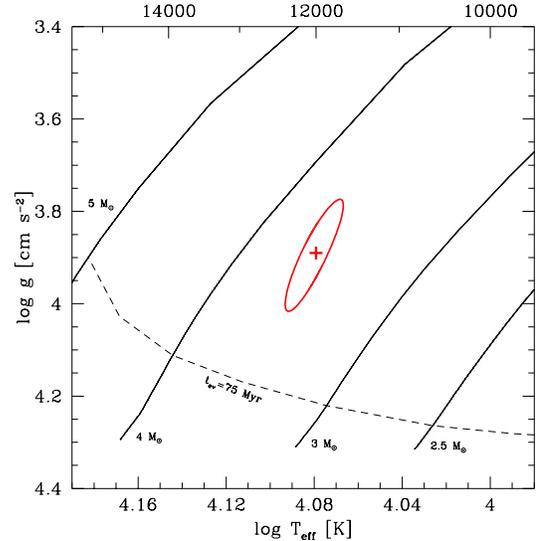}
 \caption{ \label{fig:teff}
	Measured \teff, \logg\ and the 1-$\sigma$ error ellipse for HVS5 (in red)
compared to \citet{girardi02, girardi04} solar metallicity main sequence tracks for
2.5--5~\msun\ stars (solid black lines); the $t_{ev}=75$ Myr isochrone is plotted 
for reference.  HVS5 is a $3.62\pm0.11$ \msun\ star. }
 \end{figure}

	Figure~\ref{fig:teff} compares the measured \teff\ and \logg\ with the 
latest \citet{girardi02, girardi04} solar metallicity main sequence tracks.  The 
ellipse in Figure~\ref{fig:teff} is the 68.3\% (1-$\sigma$) confidence region.  
Interpolating the tracks indicates that HVS5 is a $3.62\pm0.11$~\msun\ star.  
As an illustration of the systematic uncertainty, we derive 3.58~\msun\ from 
\citet{ekstrom12} tracks with rotation, and 3.72~\msun\ from \citet{girardi02, 
girardi04} +0.2 dex super-solar tracks.  These values are consistent within our 
1-$\sigma$ uncertainty, thus the inferred mass is relatively insensitive to 
rotation and metallicity.  

	Table~\ref{tab:hvs} summarizes our stellar parameters for HVS5.  We also 
list the parameters for HVS7, HVS8, and HVS1 measured by \citet{przybilla08b}, 
\citet{lopezmorales08}, and \citet{heber08b}, respectively.  HVS3, a probable blue 
straggler \citep{brown10b}, is not directly comparable and is not included in our 
discussion.

\begin{figure}		
 \plotone{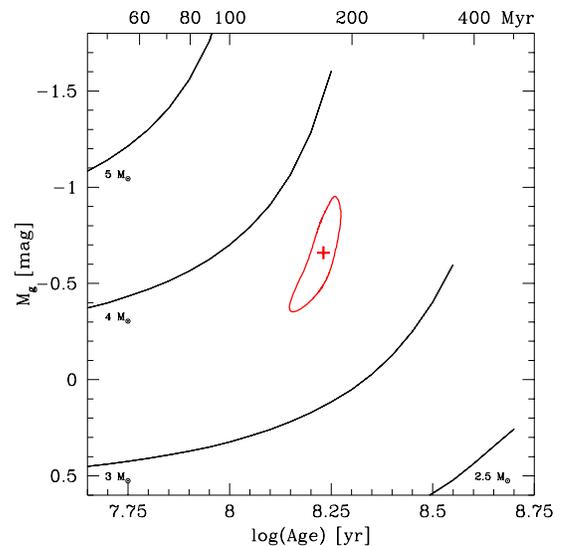}
 \caption{ \label{fig:age}
	Same tracks as Figure~\ref{fig:teff} but now plotting age versus absolute 
magnitude.  HVS5 has a formal age of $170\pm17$(stat)$\pm$30(sys)~Myr.}
 \end{figure}

\section{DISCUSSION}

\begin{deluxetable*}{lcccc}		
\tabletypesize{\footnotesize}
\tablewidth{0pt}
\tablecaption{Hypervelocity Star Properties\label{tab:hvs}}
\tablecolumns{5}
\tablehead{
  \colhead{} & 
  \colhead{HVS5} & 
  \colhead{HVS7} & 
  \colhead{HVS8} & 
  \colhead{HVS1}
}
	\startdata
\teff\ (K)	& 12000 $\pm$ 350  & 12000 $\pm$ 500  & 11000 $\pm$ 1000 & 11000 $\pm$  500 \\
\logg\ (cgs)	&  3.89 $\pm$ 0.13 &  3.8  $\pm$ 0.1  &  3.75 $\pm$ 0.25 &  3.74 $\pm$ 0.20 \\
\vsini\ (\kms)	&   133 $\pm$ 7    &    55 $\pm$ 2    &   260 $\pm$ 70   &   190 $\pm$ 40   \\
Mass (\msun)	&  3.62 $\pm$ 0.11 &  3.79 $\pm$ 0.09 &  3.49 $\pm$ 0.22 &  3.50 $\pm$ 0.18 \\
Age (Myr)	&   170 $\pm$ 17   &   170 $\pm$ 15   &   220 $\pm$ 25   &   220 $\pm$ 20   \\
$M_g$ (mag)	& -0.65 $\pm$ 0.19 & -0.94 $\pm$ 0.14 & -0.77 $\pm$ 0.35 & -0.80 $\pm$ 0.29 \\
$g_0$ (mag)	&17.557 $\pm$ 0.021&17.637 $\pm$ 0.019&17.939 $\pm$ 0.016&19.688 $\pm$ 0.023\\
$r_{GC}$ (kpc)	&    50 $\pm$ 5    &    53 $\pm$ 4    &    60 $\pm$ 10   &   130 $\pm$ 18   \\
$t_{GC}$ (Myr)	&    65 $\pm$ 7    &   105 $\pm$ 10   &   120 $\pm$ 20   &   175 $\pm$ 25   \\
\tarr=Age$-t_{GC}$ (Myr)
		&   105 $\pm$ 18   &    65 $\pm$ 18   &   100 $\pm$ 32   &    45 $\pm$ 32
	\enddata
 \end{deluxetable*}

\subsection{Hypervelocity Star Flight Times and Ages}

	We use the \citet{girardi02, girardi04} tracks to derive age and luminosity
from the measured \teff\ and \logg.  We note that the \teff-\logg\ error ellipse
becomes a banana shape in the age-$M_g$ plot (Figure~\ref{fig:age}) because of the
time evolution of these parameters.  Interpolating the tracks indicates that HVS5
has an age of $170\pm17$~Myr and an absolute magnitude of $M_g=-0.65\pm0.19$. 

	Knowing the luminosity of HVS5, we can calculate its distance and flight 
time.  HVS5 has an apparent de-reddened magnitude of $g=17.557\pm0.021$ and thus a 
heliocentric distance of $44\pm4$~kpc.  Assuming the Sun is located 8~kpc from 
the Galactic center, HVS5 has a Galactocentric distance of $r_{GC}=50\pm5$~kpc.  We 
then take the Galactic potential model of \citet{kenyon08} and calculate all 
possible trajectories that HVS5 could have given its present distance and radial 
velocity. The trajectory that passes through the Galactic center has a flight time 
of $t_{GC}=65\pm7$~Myr.  The error comes from propagating the distance and radial 
velocity errors through the trajectory calculation.

	Our Galactic center flight time estimate is appropriate for a wide range of
Milky Way starting locations because HVS5 is located at high Galactic latitude and
at large distance.  Moving the assumed starting location of HVS5 from $r_{GC}=0$~kpc
to $r_{GC}=10$~kpc changes the flight times by $\pm$8.5~Myr, which is similar to the
estimated uncertainty in flight time.

	For self-consistency, we also derive the ages and flight times of HVS7, 
HVS8, and HVS1 using the same tracks and methodology.  Table~\ref{tab:hvs} 
summarizes the derived values.  The HVSs have ages of 170--220~Myr and flight times 
of 45--175~Myr.

\subsection{Links to Unbound Ejection Processes}

	There are many ways to eject stars from their place of origin, but few
processes can accelerate stars to unbound velocities.  Because most B stars are
binaries \citep[e.g.][]{chini12}, disk ``runaway'' B stars are explained by binary
disruption mechanisms.  In the case of a supernova in a binary system, the timescale
of the process is the lifetime of the $\gtrsim$10~\msun\ star that explodes,
$10^6$--$10^7$~yr.  In the case of dynamical 3- and 4-body encounters, e.g.\ in
young star clusters, massive stars are necessary to attain the unbound velocities of
HVSs and thus the timescale of the dynamical process is also $\simeq$10$^7$~yr.  
Except in rare circumstances \citep{gualandris07, gvaramadze09, silva11}, no runaway
mechanism is expected to yield unbound velocities for 3~\msun\ stars
\citep{portegies00, perets12}.

	A more energetic and higher ejection rate process exists in the Galactic 
center:  HVSs ejected by the central black hole \citep{hills88}.  The B stars that 
presently orbit Sgr~A* on short-period, eccentric orbits are, in this scenario, the 
former companions of HVSs; their progenitors are believed to have formed further out 
and then moved in towards the black hole through dynamical processes 
\citep[e.g.][]{perets07, perets08c, madigan09, madigan11}.

	In principle, there is no upper limit to the arrival time \tarr\ for the 
central black hole ejection process. The black hole is always there, and on-going 
star formation \citep[e.g.][]{lu09} provides a constant supply of new stars.  To 
derive a typical \tarr, we consider the `loss cone,' the set of orbits which have a 
distance of closest approach within the black hole's tidal radius. For an ensemble 
of stars formed close to the black hole, a few will have orbits that interact with 
the black hole on $\lesssim1$~Myr timescales and so are quickly removed.  As a 
result of dynamical interactions with other massive objects or the long-term 
evolution of chaotic orbits within a triaxial potential, the remaining stars will
`fill' the loss cone with timescales of 100~Myr to 1~Gyr \citep{yu03, 
merritt04, wang04, perets07}. 

	Timescale thus provides an important distinction between the central black 
hole and disk runaway ejection processes.  The disk runaway scenarios must eject 
stars within the 1--10~Myr lifetimes of massive stars to attain unbound velocities. 
The central black hole can eject unbound stars at any time, however we expect that 
stars formed in the Galactic center will have typical arrival times of 0.1--1~Gyr. 
For the HVSs studied here, an upper limit is provided by their finite lifetimes.

\subsection{Comparison with Observations}

	From Table~\ref{tab:hvs}, observed HVSs have \tarr\ $\approx50$--100~Myr.  
The formal error in \tarr\ is likely an underestimate of the true error, however.  
Perhaps the best estimate of systematic error comes from comparing the measured 
stellar parameters with different sets of stellar evolution tracks.  For HVS5, the 
\citet{ekstrom12} tracks for rotating stars give a longer age of $200\pm23$~Myr, 
while the \citet{girardi02, girardi04} +0.2 dex super-solar tracks give a shorter 
age of $142\pm16$ Myr.  Taking the $\pm$30~Myr age spread as the systematic error 
rules out an ejection in the first 10~Myr of HVS5's lifetime at the 3-$\sigma$ 
level.  This confidence level is corroborated by the directly measured parameters:  
the \logg\ and \teff\ of a 75 Myr old 3.62~\msun\ star differ by 2.1-$\sigma$ and 
3.3-$\sigma$, respectively, with respect to HVS5's present values (see Figure 
\ref{fig:teff}). Thus for HVS5 we rule out a possible \tarr\ $\lesssim10$~Myr at the 
3-$\sigma$ level, an interesting and important constraint on its origin.

	The hyper-runaways first discovered by \citet{heber08} show a contrasting
result.  HD 271791 is an unbound 11~\msun\ B star. The observed proper motion shows
it was ejected in the direction of rotation from the outer disk \citep{heber08}.  
The star has an age of $25\pm5$~Myr and a flight time from the disk of $25\pm6$~Myr.  
Formally, the star has \tarr$=0\pm8$~Myr.  The marginally unbound 5~\msun\ B star
HIP 60350 is similar \citep{irrgang10}.  It has an age of $45^{+15}_{-30}$~Myr and a
flight time from the disk of $14\pm3$~Myr. Thus, \tarr$=31^{+15}_{-30}$~Myr. The
short arrival times are consistent with both the supernova ejection scenario
\citep{przybilla08c} and the dynamical ejection scenario \citep{gvaramadze09b}.  
Contrasting the derived arrival times with arrival times for HVSs underscores the
usefulness of \tarr\ as a model discriminant.

	Other objects are more ambiguous.  \citet{tillich09} discovered the 
marginally unbound 2.5~\msun\ A star J0136+2425 with a derived age of 245~Myr and a 
flight time of 12~Myr if it comes from the disk.  Accepting modern Milky Way halo 
mass estimates of $\simeq1.7\times10^{12}$~\msun\ \citep[e.g.][]{gnedin10, 
przybilla10}, it is bound to the Milky Way and may be explained as a halo star.  
The evolved sdB star J1211+1437 has a flight time that is also a small fraction of 
its progenitor's lifetime (which may be many Gyr) \citep{tillich11}.  Given the 
$\pm$140~\kms\ uncertainty in the space motion, this sdB star is also consistent 
with being bound and thus a normal halo star.

	The HVSs are significantly unbound based on radial velocity alone.  The four
HVSs with known evolutionary state discussed here have \tarr$=50$--100~Myr, times
both larger than known hyper-runaways and larger than the maximum \tarr\ expected in
the mechanisms for producing hyper-runaways. However, their \tarr\ are close to the
arrival times expected for dynamical interactions with the black hole at the
Galactic center.

\section{CONCLUSION}

	We describe Keck HIRES spectroscopy of HVS5, one of the fastest known HVSs 
with a minimum Galactic rest frame velocity of $+663\pm3$~\kms.  The observations 
reveal that HVS5 has a projected rotation of \vsini$=133\pm7$~\kms\ and is thus a 
main sequence B star.  Comparing the measured \teff\ and \logg\ with stellar 
evolution tracks indicates that HVS5 is a $3.62\pm0.11$~\msun, $170\pm17$~Myr old 
star.  Given its present distance and radial velocity, we calculate that HVS5's 
arrival time, the time between its formation and subsequent ejection, is 
\tarr$=105\pm18$(stat)$\pm$30(sys)~Myr.

	This timescale provides an interesting new constraint on the origin of 
unbound runaways and HVSs.  Runaway B stars near the disk have \tarr$=0$--30~Myr, 
consistent with disk ejection scenarios involving a supernova in a binary system or 
a dynamical event among several massive stars. The set of B-type HVSs with known 
evolutionary states, on the other hand, have \tarr$=50$--100~Myr.  This timescale is 
difficult to reconcile with any ejection mechanism requiring a massive star to 
attain unbound ejection.  The central black hole ejection scenario, however, allows 
for any \tarr.  Thus, the derived arrival times for HVSs support the black hole 
ejection model.

	Future progress requires obtaining high resolution observations of other
HVSs to constrain their age and distance.  The age distribution of HVSs has
important implications for the epochs of star formation and the growth of the
central black hole \citep{bromley12}.  Combined with future proper motion
measurements, we hope to directly constrain the full space velocity and place of
origin of the HVSs.

\acknowledgements

	This work was supported in part by the Smithsonian Institution.  J.~Cohen 
acknowledges partial support from NSF grant AST--0908139.  This research makes use 
of NASA's Astrophysics Data System Bibliographic Services. We are grateful to the 
many people who have worked to make the Keck Telescopes and their instruments a 
reality, and who operate and maintain these observatories. The authors wish to 
extend special thanks to those of Hawaiian ancestry on whose sacred mountain we are 
privileged to be guests.  Without their generous hospitality, none of the 
observations presented herein would have been possible


\end{document}